\documentclass[a4paper,onecolumn]{quantumarticle}
\pdfoutput=1 

\usepackage[numbers,sort&compress]{natbib}
\usepackage[T1]{fontenc} 
\usepackage{mathrsfs}
\usepackage{amsbsy}
\usepackage{amsfonts}
\usepackage{enumerate}
\usepackage{bbm}
\usepackage{amsthm}
\usepackage{physics}
\usepackage{hyperref}
\usepackage{ulem}
\usepackage[export]{adjustbox}
\newtheorem{definition}{Definition}

\numberwithin{equation}{section}
\renewcommand{\tr}{\operatorname{Tr}}

\renewcommand{\tilde}{\widetilde}

\renewcommand{\H}{\mathcal{H}}

\renewcommand{\d}{\dd\hspace{-0.04em}}

\newcommand{\qme}{\mathfrak{qme}}
\newcommand{\ham}{\mathfrak{ham}}
\newcommand{\Herm}{\operatorname{Herm}}
\newcommand{\swap}{\operatorname{swap}}
\newtheorem{theorem}{Theorem}[section]

\theoremstyle{remark}

\title{A canonical Hamiltonian for open quantum systems}
\author{Patrick Hayden}
\email{phayden@stanford.edu}
\author{Jonathan Sorce}
\email{jsorce@stanford.edu}
\affiliation{Stanford Institute for Theoretical Physics, Stanford University, 382 Via Pueblo, Stanford, CA 94305}

\begin{document}
\maketitle

\begin{abstract}
If an open quantum system is initially uncorrelated from its environment, then its dynamics can be written in terms of a Lindblad-form master equation. The master equation is divided into a unitary piece, represented by an effective Hamiltonian, and a dissipative piece, represented by a hermiticity-preserving superoperator; however, the division of open system dynamics into unitary and dissipative pieces is non-unique. For finite-dimensional quantum systems, we resolve this non-uniqueness by specifying a norm on the space of dissipative superoperators and defining the canonical Hamiltonian to be the one whose dissipator is minimal. We show that the canonical Hamiltonian thus defined is equivalent to the Hamiltonian initially defined by Lindblad, and that it is uniquely specified by requiring the dissipator's jump operators to be traceless, extending a uniqueness result known previously in the special case of Markovian master equations. For a system weakly coupled to its environment, we give a recursive formula for computing the canonical effective Hamiltonian to arbitrary orders in perturbation theory, which we can think of as a perturbative scheme for renormalizing the system's bare Hamiltonian.
\end{abstract}

\section{Introduction}

One of the fundamental postulates of quantum mechanics is that the time evolution of a closed quantum system is governed by a Hamiltonian. If $\rho$ is a density operator for a closed system with Hamiltonian $H$, then the time evolution of $\rho$ is governed by the differential equation\footnote{Here and throughout the paper we use units with $\hbar = 1.$}
\begin{equation}
	\frac{\d\rho}{\d t} = - i \left[ H, \rho \right].
\end{equation}
Integrating this equation with respect to time gives rise to a unitary time evolution $\rho \mapsto U(t) \rho U(t)^{\dagger}.$

If a system with Hilbert space $\H$ can be split into subsystems as $\H = \H_A \otimes \H_B$, however, then the states of subsystems $A$ and $B$ do not necessarily evolve unitarily. If the Hamiltonian on $\H$ contains terms that couple the subsystems, then information is exchanged between the subsystems and neither one, considered as its own quantum system, undergoes unitary evolution. The subsystems $A$ and $B$ are called \emph{open quantum systems}.

Now, let $\H$ be a finite-dimensional Hilbert space, and let $\mathcal{B}(\H)$ denote the space of linear operators on $\H.$ A \textit{superoperator} on $\H$ is a linear map from $\mathcal{B}(\H)$ to itself. Under the assumptions that $\H$ is finite dimensional and that $\rho_{AB}$ at $t=0$ is a product state, one can show that there is an interval of time starting at $t=0$ for which the differential equation governing $\rho_A$ is linear in $\rho_A$. I.e., there exists an interval $[0, t_0)$ and a (possibly time-dependent) superoperator $\mathcal{L}_t : \mathcal{B}(\H_A) \rightarrow \mathcal{B}(\H_A)$ satisfying\footnote{Results of this type have appeared in several places in the literature. The most general result of which we are aware appeared first in \cite{buvzek1998reconstruction} and is explained nicely in \cite{breuer2012foundations}. Some authors take the existence of an equation like \eqref{eq:proto-qme} as a fundamental postulate of open quantum systems; it follows, for example, from the assumption that the dynamics are governed by a quantum dynamical semigroup \cite{lindblad1976generators}.}
\begin{equation} \label{eq:proto-qme}
	\frac{\d \rho_A}{\d t}
		= \mathcal{L}_t(\rho_A) \quad \text{for} \quad t \in [0, t_0).
\end{equation}
The superoperator in this expression will generically depend on the $t=0$ state of the $B$ system. It is \emph{hermiticity-preserving and trace-annihilating} (HPTA), i.e., it satisfies\footnote{The additional assumption that must be made to guarantee a \emph{Markovian} master equation, as formulated by Lindblad in \cite{lindblad1976generators}, is
\begin{equation*}
	(\mathcal{L}_t^* \otimes I_n)(M^{\dagger} M) \geq (\mathcal{L}_t^* \otimes I_n)(M^{\dagger}) M + M^{\dagger} (\mathcal{L}_t^* \otimes I_n)(M),
\end{equation*}
for any integer $n$ and any operator $M$ on $\H_A \otimes \mathbb{C}^n$, where $\mathcal{L}_{t}^*$ is the adjoint of the superoperator $\mathcal{L}_{t}$. Lindblad called HPTA superoperators satisfying this property \emph{completely dissipative}, and showed that this property follows from the assumption that the Heisenberg-picture dynamics are governed by a quantum dynamical semigroup.}
\begin{align}
	\mathcal{L}_{t}(M)^{\dagger}
		& = \mathcal{L}_t(M^{\dagger}), \\
	\tr(\mathcal{L}_{t}(M))
		& = 0.
\end{align}

For any choice of Hermitian operator $H_A(t)$, one can define a new superoperator  $\mathcal{D}_t(\rho_A) \equiv \mathcal{L}_t(\rho_A) + i [H_A(t), \rho_A]$ to obtain the equation
\begin{equation} \label{eq:qme}
	\frac{\d \rho_A}{\d t}
	= - i [H_A(t), \rho_A] + \mathcal{D}_t(\rho_A).
\end{equation}
Under such a choice, the first term in the above expression is called the \emph{Hamiltonian piece} of the open system dynamics, while the second term is called the \emph{dissipative piece}; the whole equation is called a \emph{quantum master equation}. Thus far, the choice of Hamiltonian is completely arbitrary. It is common to choose a Hamiltonian that puts the dissipator $\mathcal{D}_t$ in \emph{Lindblad form}: one can show \cite{lindblad1976generators, gorini1976completely, breuer2012foundations} that there always exists a set of ``jump operators'' $L_j$ and real coefficients $\gamma_j$ for which $\mathcal{D}_t$ takes the form
\begin{equation} \label{eq:Lindblad-dissipator}
	\mathcal{D}_t(\rho_A)
		= \sum_j \gamma_j \left[ L_j \rho_A L_j^{\dagger}
			- \frac{1}{2} \left\{ L_j^{\dagger} L_j, \rho_A \right\} \right].
\end{equation}
For completely generic dynamics, the coefficients $\gamma_j$ and the jump operators $L_j$ are time-dependent. Under the \emph{Markovian assumption} --- which is essentially that time evolution does not produce significant amounts of entanglement between the $A$ and $B$ systems --- the coefficients $\gamma_j$ can be made positive. An additional assumption of time independence is often made to make $\gamma_j$ and $L_j$ constant.

Even if we restrict to master equations with dissipators that can be written in Lindblad form, however, there is still a serious ambiguity in how the master equation \eqref{eq:qme} is divided into a Hamiltonian piece and a dissipative piece. In fact, one can show (cf. section \ref{subsec:lindblad-form}) that \emph{any} dissipator can be written in the general non-Markovian Lindblad form. In the case of Markovian dynamics, the requirement that the dissipator be written in Lindblad form is a little more restrictive, but there are still ambiguities. For example, the joint transformation
\begin{align}
	L_j
		& \mapsto L_j - \alpha_j(t) I_A, \label{eq:jump-transformation} \\
	H_A
		& \mapsto H_A + \sum_j \frac{\gamma_j}{2 i} (\alpha_j L_j^{\dagger} - \bar{\alpha}_j L_j), \label{eq:ham-transformation}
\end{align}
changes the Hamiltonian but leaves the combination $- i [H_A(t), \rho_A] + \mathcal{D}_{t}(\rho_A)$ unchanged.\footnote{This is \emph{not} the trivial ambiguity that allows one to add trace terms to any Hamiltonian without changing the dynamics --- the transformation given in equations \eqref{eq:jump-transformation} and \eqref{eq:ham-transformation} changes the Hamiltonian by terms that are not, generically, pure-trace.} This ambiguity is often resolved by requiring the jump operators $L_j$ to be traceless, as it was shown in \cite{gorini1976completely} that for Markovian master equations this condition picks out a unique dissipator. To our knowledge, however, this choice is motivated by intuition rather than by a universal, rigorous principle.

The purpose of this paper is to define a physical principle for choosing an effective Hamiltonian for the equation \eqref{eq:qme}. We do this by introducing a norm on the space of superoperators on $\H_A$ that captures how large the superoperator is in an averaged sense over the whole Hilbert space. The canonical Hamiltonian for the master equation \eqref{eq:qme} is chosen by minimizing the size of the dissipator $\mathcal{D}_t$ with respect to our chosen norm. Because the norm we choose is compatible with a certain inner product on the space of superoperators, the canonical Hamiltonian can be obtained by an orthogonal projection in the space of superoperators --- one starts with an equation of the form \eqref{eq:proto-qme}, then orthogonally projects the superoperator $\mathcal{L}_{t}$ down to the space of ``Hamiltonian superoperators'' that act by commutation with a Hermitian operator. To be precise, the canonical dissipator will be minimal with respect to the norm described in the following two definitions.

\begin{definition}
Let $\ket{0} \in \mathcal{H}$ be an arbitrary reference state in a finite-dimensional Hilbert space $\H$, and $dU$ the Haar measure over the unitary group $U(\mathcal{H}).$ The state $\ket{\psi} \equiv U \ket{0}$ is called a \textbf{Haar-random state}.
\end{definition}

\begin{definition}
Let $\ket{\psi}$ and $\ket{\phi}$ be Haar-random states in a finite-dimensional Hilbert space $\H$, and $\mathcal{D}_{t}$ a hermiticity-preserving superoperator on $\H$. The \textbf{average norm} of $\mathcal{D}_{t}$ is defined by\footnote{This quantity does not depend on the choice of reference state $\ket{0}$, which we verify by explicit calculation in section \ref{sec:main}.}
\begin{equation} \label{eq:avg-norm}
	(\lVert \mathcal{D}_t \rVert_{\text{avg}})^2
		= \overline{\bra{\psi} \overline{\mathcal{D}_t(\ketbra{\phi}{\phi})^2} \ket{\psi}},
\end{equation}
where overlines denote averages with respect to the Haar measure.
\end{definition}
\noindent We explain why we think this is a good choice of norm in section \ref{subsec:avg-norm}. We also give several other natural, inequivalent norms that all minimize on the same dissipator, which we view as further justification for this dissipator being canonical.

We will find that choosing the canonical Hamiltonian for a master equation \emph{automatically} puts the dissipator in Lindblad form with traceless jump operators. Furthermore, we will find it is the \emph{unique} dissipator with this property. This uniqueness was shown for Markovian dissipators in \cite{gorini1976completely}; we establish it for non-Markovian dissipators as well. This uniqueness is practically useful, because if one is able to write down a Lindblad-form dissipator with traceless jump operators, then one knows immediately that it is the unique, canonical dissipator. In particular, in the Markovian case, our canonical dissipator must agree with the the standard dissipator from \cite{gorini1976completely}.
	
The major claims of the previous two paragraphs are summarized in the following theorem, with proofs given in section \ref{sec:main}.
\begin{theorem}
	Let $\mathcal{H}$ be a finite-dimensional Hilbert space. For any HPTA superoperator $\mathcal{L}_{t}$, there is a unique superoperator $\mathcal{D}_{t}$ such that:
	\begin{enumerate}[(1)]
		\item There is a Hermitian operator $H$ satisfying
			\begin{equation}
				\mathcal{L}_{t}(\cdot) = - i [H, \cdot] + \mathcal{D}_{t}(\cdot).
			\end{equation}
		\item We have 
			\begin{equation}
				\lVert \mathcal{D}_{t} \rVert_{\text{avg}} \leq \lVert \tilde{\mathcal{D}}_{t} \rVert_{\text{avg}}
			\end{equation}
			for any superoperator $\tilde{\mathcal{D}_{t}}$ satisfying property (1).
	\end{enumerate}

	Furthermore, this superoperator $\mathcal{D}_{t}$ can be written in Lindblad form as in equation \eqref{eq:Lindblad-dissipator} with $\tr(L_j) = 0,$ and is the unique dissipator with this property.
\end{theorem}	
\noindent As emphasized in preceding paragraphs, the uniqueness of the traceless-jump-operator dissipator was previously established for Markovian $\mathcal{L}_{t}$ in \cite{gorini1976completely}; our theorem extends that result to the non-Markovian case.

As a final application of our methods, we will give an algorithm for computing the canonical effective Hamiltonian, and the associated dissipator, for a weakly-coupled system with time-independent Hamiltonian\footnote{We use the notation ``bare'' here because we have already used the notation $H_A$ to refer to the effective Hamiltonian for the open system dynamics on $A$; this will only equal $H_{A, \text{bare}}$ for $\lambda=0$.} $H_{AB} = H_{A, \text{bare}} + H_{B, \text{bare}} + \lambda H_{\text{interaction}}$. The algorithm allows one to compute the canonical effective Hamiltonian recursively to arbitrary orders in $\lambda$-perturbation theory. This can be thought of as a perturbative renormalization scheme for computing the effective Hamiltonian of a system that is weakly coupled to its environment.

Perhaps our main result --- that the minimal dissipator with respect to the norm \eqref{eq:avg-norm} is of the form originally written down in \cite{lindblad1976generators, gorini1976completely} --- shouldn't be so surprising. After all, in \cite{lindblad1976generators}, Lindblad defined the canonical dissipator in terms of a Haar average over the unitary group. Writing out the dissipator defined in Lindblad's proposition 5 explicitly --- and rewriting Haar integrals of the adjoint $\mathcal{L}_t^*$ in terms of $\mathcal{L}_{t}$ in a way we describe in section \ref{subsec:no-adjoint} --- gives
\begin{equation} \label{eq:Lindblad-Haar}
	\mathcal{D}^{\text{Lindblad}}_{t}(\rho_A)
		= \mathcal{L}_{t}(\rho)
			+ \frac{1}{2} \int \dd U\, \left[ U^{\dagger} \mathcal{L}_t(U) - \mathcal{L}_t(U^{\dagger}) U, \rho_A \right],
\end{equation}
or equivalently
\begin{equation} \label{eq:Lindblad-ham}
	H^{\text{Lindblad}}_{A}
		= \frac{1}{2 i} \int \dd U\, \left( U^{\dagger}
		\mathcal{L}_t(U) - \mathcal{L}_t(U^\dagger) U  \right).
\end{equation}
To our knowledge, however, the connection between equation \eqref{eq:Lindblad-Haar} and the minimization of \eqref{eq:avg-norm} has not been made in the literature. This may be because the norm in \eqref{eq:avg-norm} involves a four-moment Haar integral, while Lindblad's equation \eqref{eq:Lindblad-Haar} contains only two-moment Haar integrals, making the connection mathematically nontrivial. It may also be because modern approaches to deriving the Lindblad form of a master equation --- as presented in e.g. \cite{breuer2002theory} --- tend to follow the approach to master equations formulated in \cite{gorini1976completely}, which does not make use of Haar integration. Our goal in the present work is to establish clearly that equation \eqref{eq:Lindblad-Haar} is the result of an orthogonal projection by an inner product that induces the norm \eqref{eq:avg-norm}.

The plan of the paper is as follows.

In section \ref{sec:review}, we review the modern derivation of the Lindblad form of the dissipator \eqref{eq:Lindblad-dissipator} for initially uncorrelated quantum systems undergoing general, non-Markovian time evolution. We show how any dissipator can be written in non-Markovian Lindblad form, emphasizing the Hamiltonian ambiguity. We then show how adding the Markovian assumption lets one choose a Lindblad-form dissipator whose coefficients $\gamma_j$ are positive. To emphasize the utility of Haar integration in studying quantum master equations, the presentation we give of the Markovian case is closer to Lindblad's presentation in \cite{lindblad1976generators} than to the modern treatment in e.g. \cite{breuer2002theory}. This review section is non-essential to the logical flow of the paper, and can be skipped by eager readers provided that they are willing to take equation \eqref{eq:Lindblad-ham} on faith.

In section \ref{sec:main}, we introduce a norm on the space of superoperators that captures the size of a superoperator in an averaged sense over all of Hilbert space (cf. equation \eqref{eq:avg-norm}). We show that this norm is induced by an inner product on the space of superoperators, and use orthogonal projection to derive a formula for the dissipator that minimizes our norm, calling the corresponding Hamiltonian ``canonical.'' We show the equivalence of our canonical Hamiltonian and dissipator to those that commonly appear in the literature, both in terms of Lindblad's original equations \eqref{eq:Lindblad-Haar}, \eqref{eq:Lindblad-ham} and in the modern language of traceless jump operators. In particular, we show not only that the canonical dissipator is Lindblad form with traceless jump operators, but that it is the only dissipator with these two properties; the upshot is that if you have a quantum master equation whose dissipator is Lindblad form with traceless jump operators, then it is guaranteed to be the canonical one.

In section \ref{sec:perturbations}, we give a recursive algorithm for computing the canonical effective Hamiltonian and dissipator for a system weakly coupled to its environment, to all orders in perturbation theory; we write the canonical Hamiltonian explicitly to quadratic order in perturbation theory. This calculation was inspired by \cite{agon2018coarse}, where the authors computed an effective Hamiltonian for the infrared degrees of freedom in certain weakly interacting quantum field theories, considered as an open system with the ultraviolet degrees of freedom playing the role of the environment, to quadratic order in perturbation theory. At this order, our canonical Hamiltonian is formally identical to theirs up to the addition of corrections that naively vanish in the infinite-dimensional limit. We comment on this similarity, and on extending our definition of the canonical Hamiltonian to infinite-dimensional Hilbert spaces.

Many of the general principles of open quantum systems that we use in this paper are so well established in the field that it is difficult to decide exactly whom to cite when using them, especially since their modern derivations are so simple that we are able to present them in a self-contained fashion. Some early work on quantum dynamical semigroups can be found in \cite{kossakowski1972quantum}; Lindblad's original paper showing that generators of quantum dynamical semigroups can be written with a dissipator of the form \eqref{eq:Lindblad-dissipator} is \cite{lindblad1976generators}. A contemporary paper that established the same result in finite-dimensional systems using different techniques is \cite{gorini1976completely}. Many developments in the theory of Markovian open quantum systems followed from these original papers; a comprehensive overview can be found in \cite{breuer2002theory}. Non-Markovian master equations are studied in \cite{breuer2012foundations}.

\textbf{Note added:} After the initial posting of this work on arXiv, our choice of canonical Hamiltonian was applied by Colla and Breuer to the problem of strong-coupling quantum thermodynamics \cite{colla2021exact}. They refer to the principle we use here to pick out a canonical Hamiltonian as the ``principle of minimal dissipation,'' which we think nicely encapsulates the physical intuition behind our arguments.

\section{Principles of Markovian and non-Markovian master equations}
\label{sec:review}

This section is split into four subsections. In the first, we give a proof of equation \eqref{eq:proto-qme} for initially uncorrelated quantum systems following \cite{breuer2012foundations}. In the second, we show that equation \eqref{eq:proto-qme} can be given a Lindblad-form dissipator as in equation \eqref{eq:Lindblad-dissipator};  this result was shown in \cite{breuer2012foundations}, but our pedagogical treatment is different than the one given there. In the third, we show how the general non-Markovian Lindblad master equation simplifies under the Markovian assumption using Lindblad's original techniques from \cite{lindblad1976generators}. In the fourth, we give a Haar integration identity that allows us to rewrite Lindblad's original form of the Hamiltonian (in terms of adjoint superoperators) in the form of equation \eqref{eq:Lindblad-ham}, which will be useful in future sections.

\subsection{Time evolution by linear superoperators}
\label{subsec:superoperator-evolution}

Let $\H = \H_A \otimes \H_B$ be a finite-dimensional quantum system that evolves unitarily under the operator $U(t, t').$ If the state of the system at time $t=0$ is $\rho_A(0) \otimes \rho_B(0)$, then the state of the $A$ subsystem at time $t$ is given by
\begin{equation}
	\rho_A(t)
		= \tr_B\left[ U_{AB}(t, 0) (\rho_A(0) \otimes \rho_B(0)) U_{AB}(t, 0)^{\dagger} \right].
\end{equation}
The right-hand side of this expression is linear in $\rho_A(0)$, i.e., it is a time-dependent superoperator on $\mathcal{B}(\H_A).$ Labeling this superoperator $\mathcal{N}_t$, we may write
\begin{equation} \label{eq:finite-evolution-superoperator}
	\rho_A(t)
		= \mathcal{N}_t(\rho_A(0)).
\end{equation}
Keep in mind that the superoperator $\mathcal{N}_{t}$ will generically depend on the initial state of the $B$ system --- the above equation will not accurately reproduce the $A$-system time evolution of the state $\rho_A(0) \otimes \sigma_B(0).$

The superoperator $\mathcal{N}_{t}$, being a linear map from $\mathcal{B}(\H_A)$ to itself, has a determinant. Since $\mathcal{H}_{A}$ is finite-dimensional, we know that $\mathcal{B}(\H_A)$ is finite dimensional, and so the determinant of $\mathcal{N}_{t}$ is a polynomial in its matrix entries for any basis of operators on $\mathcal{B}(\H_A).$  The only $t$-dependence in the matrix elements of $\mathcal{N}_t$ enters through the propagator $U_{AB}(t, 0)$, which is analytic in $t$ since it is generated by a finite-dimensional Hamiltonian. So all the matrix elements of $\mathcal{N}_t$ are analytic functions of $t$, which means that $\det(\mathcal{N}_t)$ is an analytic function of $t$, which means that it vanishes either everywhere or at an isolated set of values of $t$. At $t=0$, the superoperator $\mathcal{N}_t$ acts as the identity, so we have $\det(\mathcal{N}_{0}) \neq 0.$ It follows that $\mathcal{N}_{t}$ has nonzero determinant, and is thus invertible, for all $t$ except some isolated set of singularities.

Because the superoperator $\mathcal{N}_{t}$ is invertible at $t=0$, and it only fails to be invertible on a set of isolated points, there must be some interval $[0, t_0)$ on which $\mathcal{N}_{t}$ is invertible.\footnote{We stress that $\mathcal{N}_{t}$ is invertible \emph{as a superoperator}; its inverse will not generally be a quantum channel, even though $\mathcal{N}_{t}$ is.} Taking a time derivative of equation \eqref{eq:finite-evolution-superoperator} gives
\begin{equation}
	\frac{\d \rho_A}{\dd t}
		= \dot{\mathcal{N}}_t(\rho_A(0)),
\end{equation}
where the overdot indicates the time derivative of the superoperator. Plugging in $\rho_{A}(0) = \mathcal{N}_{t}^{-1} \rho_A(t)$ gives
\begin{equation} \label{eq:derived-proto-qme}
	\frac{\d \rho_A}{\d t}
		= \mathcal{L}_t(\rho_A(t)) \quad \text{for} \quad t \in [0, t_0).
\end{equation}
with $\mathcal{L}_{t} \equiv \dot{\mathcal{N}}_{t} \circ \mathcal{N}_{t}^{-1}.$

\subsection{The Lindblad form}
\label{subsec:lindblad-form}

It is well known that any completely positive superoperator admits a Kraus decomposition \cite{kraus1971general}. I.e., if $\Lambda : \mathcal{B}(\H) \rightarrow \mathcal{B}(\H)$ is a completely positive, linear map, then there is a set of operators $E_j \in \mathcal{B}(\H)$ satisfying
\begin{equation} \label{eq:Kraus-decomposition}
	\Lambda(\rho) = \sum_{j} E_j \rho E_j^{\dagger}.
\end{equation}
Less well known is that any \emph{hermiticity-preserving}\footnote{$\mathcal{L}$ is said to be hermiticity-preserving if it satisfies $\mathcal{L}(M)^{\dagger} = \mathcal{L}(M^{\dagger})$. By splitting a generic operator $A$ into its Hermitian and anti-Hermitian parts, one can show that this is equivalent to the identity $\mathcal{L}(H)^{\dagger} = \mathcal{L}(H)$ for any Hermitian operator $H$.} superoperator $\mathcal{L}$ admits what we will call a \emph{pseudo-Kraus} decomposition
\begin{equation} \label{eq:pseudo-Kraus}
	\mathcal{L}(\rho) = \sum_{j} \gamma_j E_j \rho E_j^{\dagger},
\end{equation}
where the coefficients $\gamma_j$ are real. If the $\gamma_j$ coefficients were positive, one could redefine $E'_j \equiv \sqrt{\gamma_j} E_j$ and recover the usual Kraus decomposition \eqref{eq:Kraus-decomposition}; this is how completely positive superoperators are distinguished within the more general class of hermiticity-preserving superoperators at the level of the pseudo-Kraus decomposition.

The derivation of the pseudo-Kraus decomposition is exactly the same as the derivation of the Kraus decomposition, only diverging in the assumptions imposed at the very last step. One considers a doubled Hilbert space $\H \otimes \H$, chooses an orthonormal basis $\{\ket{\alpha}\}$ on each copy of $\H$, and defines the unnormalized maximally entangled state
\begin{equation}
	\ket{\phi}_{12} = \sum_\alpha \ket{\alpha \alpha}_{12}.
\end{equation}
One then shows that for any state $\ket{\psi} \in \H$, we have
\begin{equation} \label{eq:Kraus-intermed}
	\mathcal{L}(\ketbra{\psi}{\psi}_1)
		= \bra{\tilde{\psi}}_{2} (\mathcal{L} \otimes I)(\ketbra{\phi}{\phi}_{12}) \ket{\tilde{\psi}}_{2}
\end{equation}
with
\begin{equation}
	\ket{\tilde{\psi}}_2 = \bra{\psi}_{1} \ket{\phi}_{12}.
\end{equation}
The hermiticity-preserving property of $\mathcal{L}$ implies that $(\mathcal{L} \otimes I)(\ketbra{\phi}{\phi}_{12})$ is Hermitian, and thus has a spectral decomposition
\begin{equation}
	(\mathcal{L} \otimes I)(\ketbra{\phi}{\phi}_{12})
		= \sum_{j} \gamma_j \ketbra{j}_{12}.
\end{equation}
Plugging this into equation \eqref{eq:Kraus-intermed} and defining $E_{j}$ as the unique operator satisfying $E_{j} \ket{\psi}_1 = \bra{\tilde{\psi}}_2 \ket{j}_{12}$ gives
\begin{equation}
	\mathcal{L}(\ketbra{\psi}{\psi})
		= \sum_{j} \gamma_j E_{j} \ketbra{\psi}{\psi} E_{j}^{\dagger},
\end{equation}
which reproduces equation \eqref{eq:pseudo-Kraus} by linearity. If $\mathcal{L}$ were completely positive in addition to being hermiticity-preserving, then the coefficients $\gamma_j$ would be positive, recovering the usual Kraus decomposition for completely positive maps.

The superoperator $\mathcal{L}_{t}$ appearing in equation \eqref{eq:derived-proto-qme} is certainly hermiticity-preserving, being the composition of two hermiticity-preserving maps $\dot{\mathcal{N}}_t$ and $\mathcal{N}_{t}^{-1}.$ So there exists a pseudo-Kraus decomposition for $\mathcal{L}_{t}.$ Furthermore, $\mathcal{L}_{t}$ has the property that it is \emph{trace-annihilating}, i.e., its adjoint annihilates the identity, which is equivalent to the operator equation
\begin{equation} \label{eq:trace-destroying-Kraus}
	\sum_j \gamma_j E_j^{\dagger} E_j = 0.
\end{equation}
These two facts together imply
\begin{equation}
	\mathcal{L}_{t}(\rho_A)
		= \sum_{j} \gamma_j \left[ E_j \rho_A E_j^{\dagger} - \frac{1}{2} \left\{ E_j^{\dagger} E_j, \rho_A \right\} \right].
\end{equation}
which is a Lindblad form for $\mathcal{L}_{t}$ with no Hamiltonian piece. In other words, \emph{any hermiticity-preserving, trace-annihilating (HPTA) superoperator can be written as a dissipator in Lindblad form just by taking the pseudo-Kraus decomposition}. The generator of open system dynamics is just a special example.

In particular, for \emph{any} Hermitian choice of Hamiltonian $H_A$, the dissipator $\mathcal{D}_{t}(\rho_A) = i [H_A, \rho_A] + \mathcal{L}_{t}(\rho_A)$ is HPTA --- since both $\mathcal{L}_{t}$ and the Hamiltonian commutator are HPTA --- and can thus be written in Lindblad form. We hope that this observation emphasizes the importance of finding a canonical way to split a general master equation into Hamiltonian and dissipative pieces. For non-Markovian master equations, the effective Hamiltonian is completely unconstrained by requiring the dissipator to be in Lindblad form.

\subsection{Markovian master equations}

We will see now that in the Markovian case, one can always choose a Hamiltonian such that the dissipator is of Lindblad form with positive coefficients. This is more of a restriction on the effective Hamiltonian than in the non-Markovian case, as not every dissipator will have the positive-coefficient property.

The usual way to formulate the Markovian assumption is to assume that the time evolution superoperator defined in equation \eqref{eq:finite-evolution-superoperator} has the structure of a one-parameter semigroup, i.e., that it satisfies
\begin{equation}
	\mathcal{N}_{t_1 + t_2} = \mathcal{N}_{t_1} \circ \mathcal{N}_{t_2}.
\end{equation}
In \cite{lindblad1976generators}, Lindblad showed that this implies that the adjoint of the superoperator $\mathcal{L}_{t}$ generating the time evolution satisfies\footnote{The adjoint $\mathcal{L}^*$ of a superoperator $\mathcal{L}$ is defined by the equation \begin{equation*}\tr(\mathcal{L}^*(A)^{\dagger} B) = \tr(A^{\dagger} \mathcal{L}(B)).\end{equation*}}
\begin{equation} \label{eq:Markovian-postulate}
	(\mathcal{L}_t^* \otimes I_n)(M^{\dagger} M)
		\geq (\mathcal{L}_t^* \otimes I_n)(M^{\dagger}) M + M^{\dagger} (\mathcal{L}_t^* \otimes I_n)(M),
\end{equation}
for any integer $n$ and any operator $M$ on $\H_A \otimes \mathbb{C}^n$.\footnote{In fact, the same equation holds for $\mathcal{L}_{t}$, and one could choose to work entirely in the Schr\"{o}dinger picture without making reference to the adjoint. This would lead to a slightly different form of the traditional Lindblad equation \eqref{eq:Lindblad-dissipator}, where the term $L_j^{\dagger} L_j$ is replaced by $L_j L_j^{\dagger}$.} Actually, the assumption that the dynamics are governed by a quantum dynamical semigroup is too strong, as it fixes the superoperator $\mathcal{L}_{t}$ to be time-independent. For this reason, we will take equation \eqref{eq:Markovian-postulate} to be the fundamental postulate of time-dependent Markovian dynamics.

We will let $\d U$ be the Haar measure on the group of unitary matrices on $\H_A$, normalized to satisfy $\int \d U = 1.$ Define a superoperator $\Psi_t$ by
\begin{equation} \label{eq:Psi-superoperator}
	\Psi_t(M)
		= \int \d U\, 
			\left[ \mathcal{L}_t^*(M)
					- \mathcal{L}_t^*(U^{\dagger}) U M
					- M^{\dagger} U^{\dagger} \mathcal{L}_t^*(U) \right].
\end{equation}
Now, let $n$ be any integer and $P$ any positive operator on $\H_A \otimes \mathbb{C}^n.$ Since $P$ is a positive operator, it has a square root $X$ and can be written $P = X^{\dagger} X.$ We have
\begin{equation}
	\Psi_t(P)
		= \int \d U\, 
			\left[ \mathcal{L}_t^*(X^{\dagger} X)
				- \mathcal{L}_t^*(U^{\dagger}) U X^{\dagger} X
				- X^{\dagger} X U^{\dagger} \mathcal{L}^*_t(U) \right].
\end{equation}
We now use a trick due to Lindblad. Let $\{V_j\}$ be a set of unitary operators that forms a basis for $\mathcal{B}(\H_A\otimes \mathbb{C}^n).$ We can expand $X$ as $X = \sum_{j} c_j V_j.$ But we can exploit the fact that the Haar measure is invariant under unitary multiplication to make, e.g., the substitution $U \mapsto U V_j$, giving
\begin{align}
	\int \d U\, \mathcal{L}_t^*(U^{\dagger}) U X^{\dagger} X
		& = \sum_j c_j^* \int \d U\, \mathcal{L}_t^*(U^{\dagger}) U V_j^{\dagger} X \nonumber \\
		& = \sum_j c_j^* \int \d U\, \mathcal{L}_t^*(V_j^{\dagger} U^{\dagger}) U X \nonumber \\
		& = \int \d U\, \mathcal{L}_{t}^*(X^{\dagger} U^{\dagger}) U X,
\end{align}
and similarly
\begin{equation}
	\int \d U\, X^{\dagger} X U^{\dagger} \mathcal{L}_t^*(U)
		= \int \d U\, X^{\dagger} U^{\dagger} \mathcal{L}_t^*(U X).
\end{equation}
Putting these equations together gives
\begin{equation}
	\Psi_t(P)
		= \int \d U\, 
			\left[ \mathcal{L}_t^*(X^{\dagger} U^{\dagger} U X)
				- \mathcal{L}_t^*(X^{\dagger} U^{\dagger}) U X
				- X^{\dagger} U^{\dagger} \mathcal{L}_t^*(U X) \right].
\end{equation}
Applying the Markovian assumption \eqref{eq:Markovian-postulate} with $M = U X$ tells us that the integrand is positive, and thus that $\Psi_t$ is a completely positive superoperator.

Now, using the definition of $\Psi_t$ in equation \eqref{eq:Psi-superoperator} and the trace-annihilating property $\tr(\mathcal{L}_{t}(A)) = 0$ --- which is equivalent to $\mathcal{L}_{t}^*(I) = 0$ --- it is straightforward to verify the equation
\begin{equation} \label{eq:Lindblads-full-Haar-equation}
	\mathcal{L}_{t}^*(\rho_A)
		= i \left[ \frac{1}{2i} \int \d U\, \left(\mathcal{L}_{t}^*(U^{\dagger}) U
				- U^{\dagger} \mathcal{L}_{t}^*(U) \right), \rho_A \right]
			+ \Psi_t(\rho) - \frac{1}{2} \left\{ \Psi_t(I), \rho_A \right\}.
\end{equation}
The operator in the first entry of the commutator is easily verified to be Hermitian, and is one possible choice for the effective Hamiltonian $H_A.$ Writing $\Psi_t$ in terms of a Kraus decomposition $\Psi_t(\rho_A) = \sum_j \gamma_j L_j^{\dagger} \rho_A L_j$ with $\gamma_j > 0$ immediately gives the Markovian Lindblad form for the adjoint,
\begin{equation}
	\mathcal{L}_{t}^*(\rho_A)
		= i \left[ H_A, \rho_A \right]
				+ \sum_j \gamma_j \left(L_j^{\dagger} \rho L_j - \frac{1}{2} \left\{ L_j^{\dagger} L_j, \rho_A \right\} \right).
\end{equation}
Taking the adjoint of this superoperator gives
\begin{equation}
	\mathcal{L}_{t}(\rho_A)
		= - i \left[ H_A, \rho_A \right]
			+ \sum_j \gamma_j \left(L_j \rho L_j^{\dagger} - \frac{1}{2} \left\{	L_j^{\dagger} L_j, \rho_A \right\} \right).
\end{equation}

By writing down equation \eqref{eq:Lindblads-full-Haar-equation}, Lindblad made a particular choice of effective Hamiltonian and dissipator for the $A$ system. In fact, one can show that Lindblad's choice of dissipator can be expressed with traceless jump operators, making the connection between Lindblad's original work and the modern perspective outlined in our introduction. This claim will be much easier to show with the techniques we develop in the following section, so we defer the proof to section \ref{subsec:traceless-jumps}.

\subsection{Getting rid of the adjoint}
\label{subsec:no-adjoint}

The effective Hamiltonian given in the previous section, according to Lindblad's derivation, is
\begin{equation} \label{eq:adjoint-average-hamiltonian}
	H_A
		= \frac{1}{2 i}
			\int dU \left( \mathcal{L}_{t}^*(U^{\dagger}) U
					- U^{\dagger} \mathcal{L}_{t}^*(U) \right).
\end{equation}
It will be useful, in future sections, to rewrite this integral in terms of the original superoperator $\mathcal{L}_{t}$. We gave this formula in the introduction in equation \eqref{eq:Lindblad-ham}; here we explain the connection between these two equations.

The multiplicative invariance of the Haar measure implies that for any operator $M$, the integral $\int \dd U\, U^{\dagger} M U$ commutes with any other operator $N$. In particular, for a unitary operator $V$, the substitution $U \mapsto U V^{\dagger}$ gives
\begin{equation}
	\int \dd U\, U^{\dagger} M U V = V \int \dd U\, U^{\dagger} M U,
\end{equation}
with the general case following by decomposing $N$ in a basis of unitary operators as in the previous subsection.

Writing out the superoperators $\mathcal{L}_{t}$ and $\mathcal{L}_{t}^*$ in terms of their pseudo-Kraus decompositions
\begin{align}
	\mathcal{L}_{t}(M)
		& = \sum_j \gamma_j E_j M E_j^{\dagger}, \\
	\mathcal{L}_{t}^*(M)
		& = \sum_j \gamma_j E_j^{\dagger} M E_j,
\end{align}
one can rewrite the integral in \eqref{eq:adjoint-average-hamiltonian} as
\begin{equation}
	H_A
		= \frac{1}{2 i}
			\sum_j \gamma_j \int \d U\, \left( E_j^{\dagger} U^\dagger E_j U
				- U^{\dagger} E_j^{\dagger} U E_j \right).
\end{equation}
Using the commutativity property of the preceding paragraph, this integral can be rewritten as
\begin{equation}
	H_A
		= \frac{1}{2 i}
		\sum_j \gamma_j \int \d U\, \left( U^{\dagger} E_j U E_j^{\dagger} - E_j U^{\dagger} E_j^{\dagger} U \right),
\end{equation}
or, equivalently,
\begin{equation}
	H_A
	= \frac{1}{2 i}
		\int \d U\,
			\left( U^{\dagger} \mathcal{L}_t(U) - \mathcal{L}_{t}(U^{\dagger}) U \right).
\end{equation}

\section{The canonical Hamiltonian}
\label{sec:main}

This section contains our main results. In the first subsection, we define a norm on the space of quantum master equations, and sketch how orthogonal projection can be used to compute the unique Hamiltonian and dissipator that are maximal and minimal, respectively, in our chosen norm. We also comment on two other natural choices of norm that pick out the same canonically minimal dissipator, and indicate how to check this with a simple calculation. In the second subsection, we compute the orthogonal projection explicitly, and show that the canonical Hamiltonian picked out by our procedure is exactly Lindblad's Hamiltonian \eqref{eq:Lindblad-ham}. In the third subsection, we show that there is only one dissipator that can be written in Lindblad form with traceless jump operators, and that this coincides with the canonical dissipator.

\subsection{The average norm on quantum master equations}
\label{subsec:avg-norm}

Let us begin by defining, for any finite-dimensional Hilbert space $\H$, a vector space $\qme(\H)$ of quantum master equations on $\H$. As discussed in the introduction and in section \ref{subsec:lindblad-form}, any quantum master equation
\begin{equation}
	\frac{\d \rho}{\d t} = \mathcal{L}_{t}(\rho)
\end{equation}
has the property that the superoperator $\mathcal{L}_{t}$ is HPTA --- hermiticity preserving and trace annihilating. It is straightforward to verify that real-linear combinations of HPTA superoperators are HPTA, so we will define $\qme(\H)$ to be the real vector space of HPTA superoperators on $\H$. We can also define a subspace, $\ham(\H)$, of quantum master equations generated by a Hamiltonian. I.e., an HPTA superoperator $\Phi$ is in $\ham(\H)$ if and only if there exists a Hermitian operator $H$ with
\begin{equation}
	\Phi(M) = - i [H, M].
\end{equation}
In this language, a separation of $\mathcal{L}_{t}$ into a Hamiltonian piece and a dissipative piece is a way of writing one element of $\qme(\H)$ as a linear combination of two other elements of $\qme(\H)$, one of which lies in $\ham(\H).$

To define what it means for a particular choice of dissipator to be ``small,'' we must choose a norm on $\qme(\H).$ One common choice of norm on superoperators is an infinity norm, where the norm of a superoperator $\mathcal{L}$ is taken to be the largest singular value of $\mathcal{L}$ considered as an operator on $\mathcal{B}(\H)$. For superoperators generating quantum master equations, however, it is unreasonable to expect the physical dissipator to be small with respect to this norm. For example, suppose the physical system we are considering is a periodic spin chain with nearest-neighbor interactions, divided into two contiguous intervals $A$ and $B$. Degrees of freedom lying near the edges of $A$, which are very close to the $B$ system, should have strongly dissipative dynamics, while degrees of freedom lying in the interior of $A$ should have dynamics that are more Hamiltonian. A dissipator obeying this physical principle would be large in the infinity norm, because it would have large eigenvalues on states with large amplitudes at the boundary of the open system.

Instead, it seems to us most sensible to define a norm that captures how big the dissipator is in an averaged sense over all of the degrees of freedom in $A$. This way, a dissipator that is quite large on boundary degrees of freedom can still be considered small provided that it doesn't disturb the interior degrees of freedom too badly. So for a superoperator $\mathcal{L}$ in $\qme(\H)$ and a Haar-random state $\ket{\phi}$ in $\H$, we will start by considering the operator
\begin{equation}
	\mathcal{L}(\ketbra{\phi}{\phi}).
\end{equation}
Because $\mathcal{L}$ is linear, the Haar average of this operator won't actually tell us anything about the average behavior of $\mathcal{L}$ --- it will only tell us about the behavior of $\mathcal{L}$ on the average $\overline{\ketbra{\phi}{\phi}}$, which is maximally mixed. Instead, we can consider the square
\begin{equation}
	\mathcal{L}(\ketbra{\phi}{\phi})^2.
\end{equation}
Since this expression is now nonlinear in $\ketbra{\phi}{\phi}$, its Haar average is a meaningful quantifier of the average size of $\mathcal{L}.$ Finally, because $\overline{\mathcal{L}(\ketbra{\phi}{\phi})^2}$ is an operator, we can capture its average size by taking its average expectation value with respect to a second Haar-random state $\ket{\psi}.$ The final number we will associate with the average size of $\mathcal{L}$ is
\begin{equation}
	(\lVert \mathcal{L} \rVert_{\text{avg}})^2
		= \overline{\bra{\psi} \overline{\mathcal{L}(\ketbra{\phi}{\phi})^2} \ket{\psi}}.
\end{equation}

This is a norm on $\qme(\H)$ induced by the inner product
\begin{equation} \label{eqn:inner-product}
	\langle \mathcal{M}, \mathcal{N} \rangle_{\text{avg}}
		= \overline{\bra{\psi} \overline{\mathcal{M}(\ketbra{\phi}{\phi}) \mathcal{N}(\ketbra{\phi}{\phi})} \ket{\psi}}.
\end{equation}
More precisely, let $\ket{0}$ be a reference state on $\H,$ and let $U$ and $V$ be two independent Haar-random unitary operators. The Haar random states $\ket{\phi}$ and $\ket{\psi}$ are defined as $U \ket{0}$ and $V \ket{0},$ respectively, and equation \eqref{eqn:inner-product} can be written out explicitly as
\begin{equation}
	\langle \mathcal{M}, \mathcal{N} \rangle_{\text{avg}}
		= \int dU dV \bra{0} V^{\dagger} \mathcal{M}(U \ketbra{0}{0} U^{\dagger}) \mathcal{N}(U \ketbra{0}{0} U^{\dagger}) V \ket{0}.
\end{equation}	
It can be shown that this quantity is independent of the reference state $\ket{0}$; this follows from the explicit calculation of $\langle \cdot, \cdot \rangle_{\text{avg}}$ performed in the following subsection, the final result of which is given in equation \eqref{eq:tensor-network-inner-product}.

Positive definiteness of the inner product holds because $\mathcal{L}(\ketbra{\phi}{\phi})^2$ is a positive operator, so $\lVert \mathcal{L} \rVert_{\text{avg}}$ vanishes if and only if $\mathcal{L}(\ketbra{\phi}{\phi})^2$ vanishes for all states $\ket{\phi}$. Because $\mathcal{L}(\ketbra{\phi}{\phi})$ is Hermitian, its square vanishes if and only if it vanishes. It follows that $\lVert \mathcal{L} \rVert_{\text{avg}}$ vanishing is equivalent to $\mathcal{L}(\ketbra{\phi}{\phi})$ vanishing for all $\ket{\phi}$ in $\H$, i.e., to $\mathcal{L}$ vanishing as a superoperator.

With respect to this inner product, the vector space $\qme(\H)$ can be decomposed as
\begin{equation}
	\qme(\H) = \ham(\H) \oplus \ham(\H)^{\perp}.
\end{equation}
If $P$ is the orthogonal projector onto $\ham(\H)$, then for any HPTA superoperator $\mathcal{L}$, we have a canonical division into Hamiltonian and dissipative pieces as
\begin{equation}
	\mathcal{L} = P(\mathcal{L}) + [\mathcal{L} - P(\mathcal{L})].
\end{equation}
Because this division is induced by an orthogonal projection, the dissipator will have the smallest average-norm of any dissipator for all possible divisions. This is the sense in which the Hamiltonian and dissipator defined by $P$ are canonical.

In preparing this work, we considered two other natural choices of norm on $\mathfrak{qme}(\mathcal{H})$, both of which are induced by inner products. They are the Hilbert-Schmidt inner product on superoperators,
\begin{equation}
	\langle \mathcal{M}, \mathcal{N} \rangle_{\text{HS}}
		= \tr_{\mathfrak{qme}(\mathcal{H})}[\mathcal{M}^* \circ \mathcal{N}],
\end{equation}
and the Hilbert-Schmidt inner product on Choi matrices,
\begin{equation}
	\langle \mathcal{M}, \mathcal{N} \rangle_{\text{CHS}}
		= \tr_{\mathcal{H} \otimes \mathcal{H}}[\operatorname{Choi}(\mathcal{M})^{\dagger} \operatorname{Choi}(\mathcal{N})].
\end{equation}
In the first inner product, $\mathcal{M}^*$ is the adjoint of $\mathcal{M}$, and the trace is the trace of $\mathcal{M}^* \circ \mathcal{N}$ as a superoperator, i.e., as a linear operator on the space $\mathcal{H} \otimes \mathcal{H}^*.$ In the second inner product, $\operatorname{Choi}(\mathcal{N})$ is the Choi matrix corresponding to the superoperator $\mathcal{M}$, defined as in \cite{choi1975completely}. Somewhat amazingly, one can show that all three inner products --- $\langle \cdot, \cdot \rangle_{\text{avg}},$ $\langle \cdot, \cdot \rangle_{\text{HS}},$ and $\langle \cdot, \cdot \rangle_{\text{CHS}}$ --- induce, for any $\mathcal{L} \in \mathfrak{qme}(\mathcal{H}),$ the same decomposition into Hamiltonian and dissipative pieces. This is because all three of the inner products agree (up to constant prefactors) when at least one entry of the inner product is in $\mathfrak{ham}(\mathcal{H}).$ We compute the inner product between an element of $\mathfrak{ham}(\mathcal{H})$ and a generic element of $\mathfrak{qme}(\mathcal{H})$ for the inner product $\langle \cdot, \cdot \rangle_{\text{avg}}$ in equation \eqref{eq:tensor-network-inner-product-ham}. It is a straightforward calculation to verify that the formula is the same, up to prefactors, for $\langle \cdot, \cdot \rangle_{\text{HS}}$ and $\langle \cdot, \cdot \rangle_{\text{CHS}}.$ As such, all three choices of inner product induce the same orthogonal projection from $\mathfrak{qme}(\mathcal{H})$ onto $\mathfrak{ham}(\mathcal{H}).$ In the remainder of the paper, we will deal only with $\langle \cdot, \cdot \rangle_{\text{avg}},$ as we find it to be the most physically motivated choice; however, we take the fact that the canonical Hamiltonian agrees for all three choices of inner product as justification for thinking of it as natural.

\subsection{Computing the projection}
\label{subsec:projection-computation}

The canonical Hamiltonian of $\mathcal{L}$ can be found by computing the superoperator
\begin{equation}
	\mathcal{L}_{\text{ham}}
		= \sum_{j} \Phi_j \langle \Phi_j, \mathcal{L} \rangle_{\text{avg}},
\end{equation}
where $\{\Phi_j\}$ is a basis of Hamiltonian superoperators that is orthonormal with respect to $\langle \cdot, \cdot \rangle_{\text{avg}}$. Once $\mathcal{L}_{\text{ham}}$ is computed, we then find a Hamiltonian $H$ with $\mathcal{L}_{\text{ham}}(\rho) = - i [H, \rho]$; as usual in quantum mechanics, this Hamiltonian will be determined only up to the addition of an overall trace term. To deal with this ambiguity, we will require that the canonical Hamiltonian be traceless, though this ``gauge choice'' will not affect the dynamics.

It will first be useful to verify that $\ham(\H)$ is isomorphic, as a vector space, to the space of traceless Hermitian operators on $\H$. For any Hermitian operator $H$, we define the superoperator $\Phi_H$ by
\begin{equation}
	\Phi_H(\rho) = - i [H, \rho].
\end{equation}
This is a linear map from $\Herm(\H)$, the space of Hermitian matrices on $\H$, into $\ham(\H).$ Surjectivity follows from the definition of $\ham(\H)$; injectivity follows from the fact that the only operators that commute with all other operators are multiples of the identity.

The next thing we need to know is how the inner product on $\qme(\H)$ is mapped, via this isomorphism, to an inner product on $\Herm(\H)$; we will need this information to construct an orthonormal basis of superoperators on $\ham(\H)$. So given two traceless Hermitian operators $\H_1$ and $\H_2$, we need to compute the quantity
\begin{equation}
	\langle \Phi_{H_1}, \Phi_{H_2} \rangle_{\text{avg}}
		= \overline{\bra{\psi} \overline{\Phi_{H_1}(\ketbra{\phi}{\phi}) 			\Phi_{H_2}(\ketbra{\phi}{\phi})} \ket{\psi}}.
\end{equation}
The Haar average over a random state $\ket{\phi}$ is defined by fixing a reference state $\ket{0}$ and defining $\ket{\phi} = U \ket{0}$ for $U$ a Haar-random unitary. So the $\ket{\psi}$ average in the above expression can be rewritten as
\begin{equation}
	\langle \Phi_{H_1}, \Phi_{H_2} \rangle_{\text{avg}}
		= \int \d U\, \bra{0} U^{\dagger} \overline{\Phi_{H_1}(\ketbra{\phi}{\phi}) \Phi_{H_2}(\ketbra{\phi}{\phi})} U \ket{0}.
\end{equation}
For any operator $M$, elementary techniques of Haar integration can be used to verify the identity\footnote{This can be verified directly by using the observation from subsection \ref{subsec:no-adjoint} that $\int \d U\, U^{\dagger} M U$ commutes with every operator, which means it must be proportional to the identity, and then fixing the constant of proportionality by taking a trace. Alternatively, one can use the general formulas for arbitrary-moment Haar integrals derived in \cite{collins2006integration}.}
\begin{equation}
	\int \d U\, U^{\dagger} M U = \frac{1}{\dim(\H)} \tr(M) I,
\end{equation}
giving
\begin{equation} \label{eq:trace-of-average}
	\langle \Phi_{H_1}, \Phi_{H_2} \rangle_{\text{avg}}
		= \frac{1}{\dim(\H)}\tr\left[ \overline{\Phi_{H_1}(\ketbra{\phi}{\phi}) 		\Phi_{H_2}(\ketbra{\phi}{\phi})}\right].
\end{equation}

Computing the four-moment Haar average over $\ket{\phi}$ states is slightly more complicated; we will find it the calculation easiest to express using tensor network diagrams. In this visual language, a tensor --- whether it be a state, operator, or superoperator --- is represented as a box with one leg for each of its tensor indices. We will mark legs corresponding to vector indices with outward-pointing arrows, and legs corresponding to dual vector indices with inward-pointing arrows. For example, the following visual equation shows a unitary operator $U$, a state $\ket{0}$, and the state $U \ket{0}.$
\begin{equation}
	\includegraphics{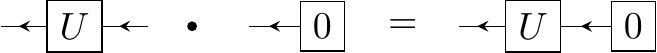}
\end{equation}
The visual equation below shows a superoperator $\mathcal{L}$ --- which must have two vector indices and two dual vector indices, since it maps operators to operators --- acting on an operator $M$ to form the operator $\mathcal{L}(M)$.
\begin{equation} \label{eq:superoperator-TN}
	\includegraphics{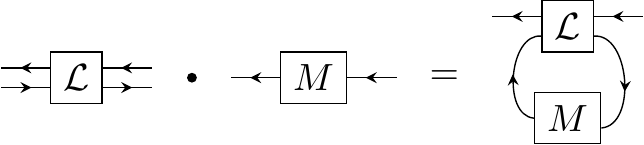}
\end{equation}
The most general four-moment Haar integral, which can be computed using the formulas of \cite{collins2006integration}, can be written as
\begin{equation}
	\int \d U\, U \otimes U^{\dagger} \otimes U \otimes U^{\dagger}
		= \frac{1}{d^2 - 1} \left( V_{(12) (34)} + V_{(14)(23)} \right)
			- \frac{1}{d(d^2 - 1)} \left( V_{(1 2 3 4)} + V_{(4 3 2 1)} \right).
\end{equation}
In this expression we have simplified notation by writing $d \equiv \dim(\H)$, and we have introduced the symbol $V_{\sigma}$ for the unitary operator that maps system $j$ to system $\sigma(j)$, where $\sigma$ is a permutation in $S_4$ --- for example, the unitary $V_{(12) (34)}$ swaps system 1 with system 2, and swaps system 3 with system 4. In tensor network notation, this general integral is drawn as follows:
\begin{equation} \label{eq:four-moment-haar-diagram}
	\includegraphics[valign=c]{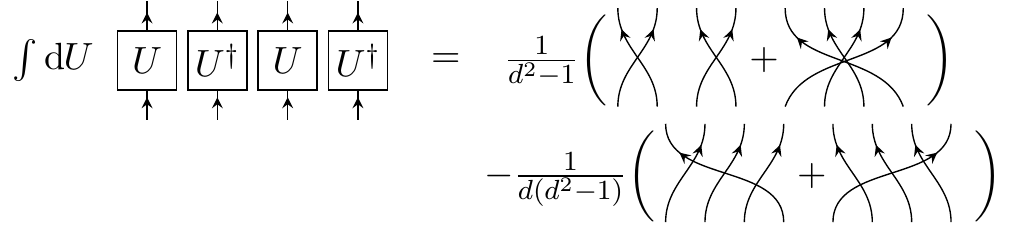}.
\end{equation}

Now, the expression we want to compute, equation \eqref{eq:trace-of-average}, can be written as
\begin{equation}
	\langle \Phi_{H_1}, \Phi_{H_2} \rangle_{\text{avg}}
		= \frac{1}{d} \int \d U\, \tr \left[ \Phi_{H_1}(U \ketbra{0}{0} U^{\dagger}) \Phi_{H_2} (U \ketbra{0}{0} U^{\dagger}) \right],
\end{equation}
and written in tensor network notation as
\begin{equation}
	\langle \Phi_{H_1}, \Phi_{H_2} \rangle_{\text{avg}}
		= \includegraphics[valign=c]{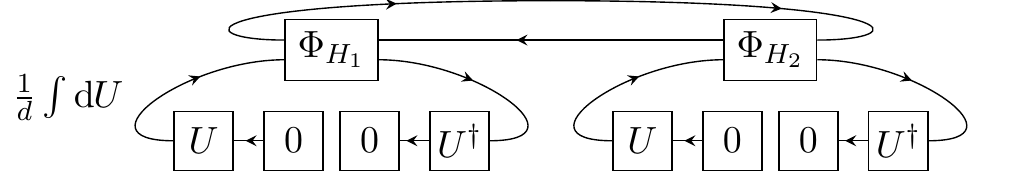}.
\end{equation}
This expression has two $U$ operators and two $U^{\dagger}$ operators, so we are free to substitute in the identity \eqref{eq:four-moment-haar-diagram} and simplify to obtain the expression
\begin{equation} \label{eq:tensor-network-inner-product}
	\langle \Phi_{H_1}, \Phi_{H_2} \rangle_{\text{avg}}
		= \includegraphics[valign=c]{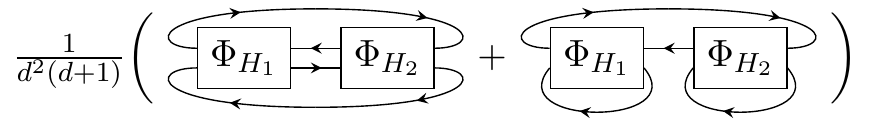}.
\end{equation}
This is a rather simple expresssion --- it expresses the inner product of superoperators in terms of a sum over particular contractions of the indices of the corresponding tensors. We can simplify it even further by putting in the explicit formula $\Phi(H_1)(\cdot) = - i [H_1, \cdot],$ which is expressed in tensor network notation as
\begin{equation}
	\includegraphics[valign=c]{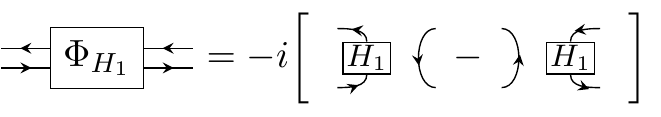}.
\end{equation}
Plugging this into equation \eqref{eq:tensor-network-inner-product} gives the expression\footnote{Remember that back in equation \eqref{eq:superoperator-TN}, chosen a convention where the ``input'' legs of a superoperator are drawn on the bottom --- neither term in the expression \eqref{eq:tensor-network-inner-product-ham} represents $\Phi_{H_2}$ acting on $H_1$, but rather a more complicated contraction of the indices of $\Phi_{H_2}$, including an insertion of $H_1.$}
\begin{equation} \label{eq:tensor-network-inner-product-ham}
	\langle \Phi_{H_1}, \Phi_{H_2} \rangle_{\text{avg}}
		= \includegraphics[valign=c]{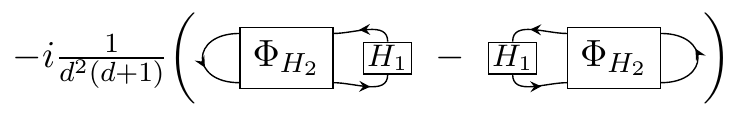}.
\end{equation}
Note that so far, we haven't actually used the assumption that $\Phi_{H_2}$ is in $\ham(\H)$; as such, we will be able to apply \eqref{eq:tensor-network-inner-product} later when considering inner products like $\langle \Phi_H, \mathcal{L} \rangle_{\text{avg}}.$ For now, though, we will put in the explicit form of $\Phi_{H_2}$ and simplify to obtain\footnote{In simplifying, we use (i) that $H_1$ and $H_2$ are traceless, and (ii) that a closed loop in a tensor network diagram represents the trace of the identity and thus contributes a factor of $d$ to the calculation.}
\begin{equation}
	\langle \Phi_{H_1}, \Phi_{H_2} \rangle_{\text{avg}}
		= \includegraphics[valign=c]{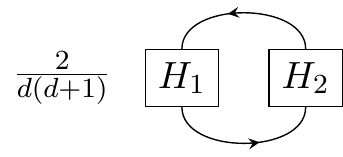}\quad.
\end{equation}
Returning momentarily to standard notation, this equation is
\begin{equation}
	\langle \Phi_{H_1}, \Phi_{H_2} \rangle_{\text{avg}}
		= \frac{2}{d (d + 1)} \tr(H_1 H_2),
\end{equation}
which is the Hilbert-Schmidt inner product up to a constant of proportionality, a result we could have anticipated from the fact that the definition of $\langle \Phi_{H_1}, \Phi_{H_2} \rangle_{\text{avg}}$ ensures its invariance under unitary conjugation of $H_1$ and $H_2$.
So a family of superoperators $\{\Phi_{H_j}\}$ is orthonormal if and only if the corresponding Hamiltonians satisfy
\begin{equation} \label{eq:Hamiltonian-orthonormality}
	\tr(H_j H_k) = \frac{d (d + 1)}{2} \delta_{jk}.
\end{equation}

Now, let $\{H_j\}$ be a basis for the real vector space of traceless Hermitian operators on $\H$ satisfying equation \eqref{eq:Hamiltonian-orthonormality}. It follows from the above discussion that $\{\Phi_{H_j}\}$ is an orthonormal basis for $\ham(\H).$ Any such basis satisfies the identity
\begin{equation} \label{eq:H-H-identity}
	\sum_j H_j \otimes H_j = \frac{d(d + 1)}{2} \left( \swap_{12} - I \right),
\end{equation}
or, in tensor network notation,
\begin{equation}
	\includegraphics[valign=c]{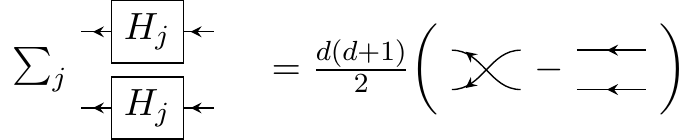}.
\end{equation}
To verify this identity, let $\ket{\alpha}$ be an orthonormal basis for $\H$, and let
\begin{align} \label{eq:M-basis}
	M_{\alpha \beta}
		& = \sqrt{\frac{d (d + 1)}{2}} \ketbra{\alpha}{\beta}
\end{align}
be a complex basis for \emph{all} operators on $\H$, normalized to satisfy
\begin{equation} \label{eq:Hilbert-Schmidt}
	\tr(M_{\alpha \beta}^{\dagger} M_{\alpha' \beta'})
		= \frac{d (d+1)}{2} \delta_{\alpha \alpha'} \delta_{\beta \beta'}.
\end{equation}
The identity
\begin{equation} \label{eq:M-swap}
	\sum_{\alpha \beta} M_{\alpha \beta} \otimes M_{\alpha \beta}^{\dagger}
		= \frac{d(d + 1)}{2} \swap_{12} 
\end{equation}
follows immediately from the definition of $M$ in equation \eqref{eq:M-basis}. Any other basis $X_j$ of operators on $\H$ satisfying
\begin{equation} \label{eq:X-orthonormality}
	\tr(X_j^{\dagger} X_k) = \frac{d (d+1)}{2} \delta_{j k}
\end{equation}
can be written as
\begin{equation}
	X_j = \sum_{\alpha \beta} u_{j, \alpha \beta} M_{\alpha \beta},
\end{equation}
where $u_{j, \alpha \beta}$ is a unitary matrix. So we have
\begin{equation}
	\sum_j X_j \otimes X_j^{\dagger}
		= \sum_{j, \alpha, \beta, \alpha', \beta'}
			u_{j, \alpha \beta} u^*_{j, \alpha', \beta'} M_{\alpha \beta} \otimes M^{\dagger}_{\alpha' \beta'}
		= \sum_{\alpha \beta} M_{\alpha \beta} \otimes M_{\alpha \beta}^{\dagger}.
\end{equation}
Hence the identity \eqref{eq:M-swap} holds not just for the special basis $M_{\alpha \beta},$ but for any basis of operators satisfying equation \eqref{eq:X-orthonormality}. Finally, we note that when $\{H_j\}$ is a basis for the real vector space of traceless Hermitian operators satisfying equation \eqref{eq:Hamiltonian-orthonormality}, $\{H_j, \sqrt{\frac{d (d+1)}{2}} I\}$ is a basis for the complex vector space of all Hermitian operators satisfying equation \eqref{eq:X-orthonormality}. From these observations, equation \eqref{eq:H-H-identity} follows.

We are finally ready to compute the canonical Hamiltonian for a generic superoperator $\mathcal{L}.$ The quantity we are trying to compute is
\begin{equation}
	\mathcal{L}_{\text{ham}}
		= \sum_{j} \Phi_{H_j} \langle \Phi_{H_j}, \mathcal{L} \rangle_{\text{avg}}.
\end{equation}
By taking advantage of the vector space structure of $\ham(\H)$, we can save ourselves a little bit of trouble and compute the canonical Hamiltonian directly as
\begin{equation}
	H
		= \sum_{j} H_j \langle \Phi_{H_j}, \mathcal{L} \rangle_{\text{avg}}.
\end{equation}
In tensor network notation, using equation \eqref{eq:tensor-network-inner-product-ham}, we can write this as
\begin{equation}
	H
		= \includegraphics[valign=c]{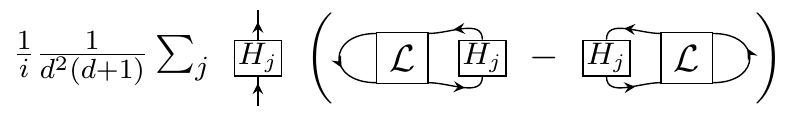}.
\end{equation}
Plugging in our identity \eqref{eq:H-H-identity} and simplifying gives
\begin{equation} \label{eq:canonical-ham-TN}
	H^{\text{canonical}}
		= \includegraphics[valign=c]{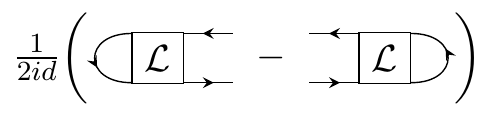}.
\end{equation}

Equation \eqref{eq:canonical-ham-TN} is an exact expression for the canonical Hamiltonian of $\mathcal{L}$ in terms of a linear combination of contractions of the corresponding tensor. All that remains in this subsection is to verify that it agrees with Lindblad's Hamiltonian; in the next subsection we will express $H^{\text{canonical}}$ in terms of jump operators. Lindblad's Hamiltonian for a quantum master equation $\mathcal{L}$ was given in equation \eqref{eq:Lindblad-ham}. It can be written in tensor network notation as
\begin{equation} \label{eq:Lindblad-ham-TN}
	H^{\text{Lindblad}}
		= \includegraphics[valign=c]{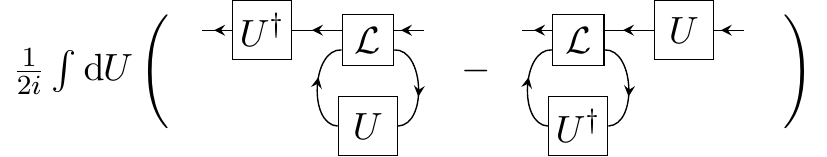}.
\end{equation}
The second-moment Haar integral is just a swap with a factor of $1/d$:
\begin{equation}
	\includegraphics[valign=c]{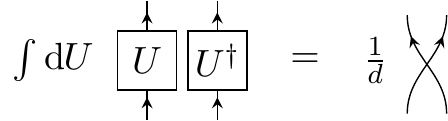}\quad.
\end{equation}
Applying this identity to equation \eqref{eq:Lindblad-ham-TN} shows that Lindblad's Hamiltonian is exactly equal to our canonical Hamiltonian \eqref{eq:canonical-ham-TN}.

\subsection{Traceless jump operators}
\label{subsec:traceless-jumps}

Let us begin by writing a generic HPTA superoperator $\mathcal{L}$ in terms of a pseudo-Kraus decomposition
\begin{equation}
	\mathcal{L}(\rho) = \sum_j \gamma_j E_j \rho E_j^{\dagger}.
\end{equation}
In tensor network notation, this is equivalent to $\mathcal{L}$ being of the form
\begin{equation}
	\includegraphics[valign=c]{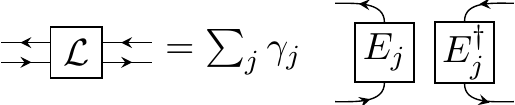}\quad.
\end{equation}
Plugging this into equation \eqref{eq:canonical-ham-TN} for the canonical Hamiltonian and simplifying gives the expression
\begin{equation}
	H^{\text{canonical}}
		= \frac{1}{2 i d} \sum_j \gamma_j \left( \tr(E_j) E_j^{\dagger} - \tr(E_j^{\dagger}) E_j \right).
\end{equation}
The canonical dissipator is given by
\begin{equation}
	\mathcal{D}^{\text{canonical}}(\rho)
		= \mathcal{L}(\rho) + i[H^{\text{canonical}}, \rho]
		= \sum_j \gamma_j
			 \left( E_j \rho E_j^{\dagger}
			 	+ \frac{1}{2 d} \left[ \tr(E_j) E_j^{\dagger} - \tr(E_j^{\dagger}) E_j, \rho \right] \right).
\end{equation}
If we use the identity $\sum_j \gamma_j E_j^{\dagger} E_j = 0$ (cf. equation \eqref{eq:trace-destroying-Kraus}), then it is straightforward to verify the equality
\begin{equation} \label{eq:canonical-dissipator-traceless}
	\mathcal{D}^{\text{canonical}}(\rho)
		= \sum_j \gamma_j \left(L_j \rho L_j^{\dagger} - \frac{1}{2} \left\{ L_j^{\dagger} L_j, \rho\right\} \right),
\end{equation}
where
\begin{equation}
	L_j
		= E_j - \frac{\tr(E_j)}{d} I
\end{equation}
is the traceless version of the pseudo-Kraus operator $E_j.$

Equation \eqref{eq:canonical-dissipator-traceless} tells us that the canonical dissipator can always be written in Lindblad form with traceless jump operators. We will now show that the canonical dissipator is the unique dissipator with this property. Suppose that there is some Hamiltonian $\tilde{H}$ and dissipator $\tilde{\mathcal{D}}$ satisfying
\begin{equation}
	\mathcal{L}(\rho) = - i [\tilde{H}, \rho] + \tilde{\mathcal{D}}(\rho),
\end{equation}
such that $\tilde{\mathcal{D}}$ can be written as
\begin{equation}
	\tilde{\mathcal{D}}(\rho)
		= \sum_j c_j \left( F_j \rho F_j^{\dagger}
			- \frac{1}{2} \left\{ F_j^{\dagger} F_j, \rho \right\} \right)
\end{equation}
for some real coefficients $c_j$ and traceless operators $F_j$. As a tensor network, this implies that $\tilde{\mathcal{D}}$ takes the form
\begin{equation}
	\includegraphics[valign=c]{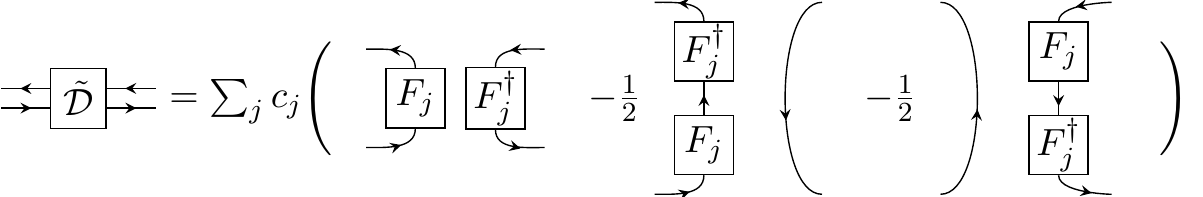}.
\end{equation}
But by plugging this into the projection formula \eqref{eq:canonical-ham-TN} and using the tracelessness of the jump operators, we see that the canonical Hamiltonian corresponding to $\tilde{\mathcal{D}}$ vanishes. In other words, the orthogonal projection of $\tilde{\mathcal{D}}$ from $\qme(\H)$ to $\ham(\H)$ vanishes. So $\mathcal{D}$ lies in the orthogonal complement $\ham(\H)^{\perp}.$ But the decomposition of $\qme(\H)$ as $\ham(\H) \oplus \ham(\H)^{\perp}$ is unique, so $\tilde{\mathcal{D}}$ must be the canonical dissipator and $\tilde{H}$ must be the canonical Hamiltonian.

As a final comment, we mention that a single dissipator may have multiple Lindblad forms. In particular, there may be multiple sets of real coefficients $\{\gamma_j\}$ and traceless operators $\{L_j\}$ whose Lindblad form give the canonical dissipator. This is important because for Markovian master equations, we know that there will always be a way of writing the canonical dissipator with \emph{positive} Lindblad coefficients and traceless jump operators. The coefficients $\{\gamma_j\}$ appearing in equation \eqref{eq:canonical-dissipator-traceless}, however, will not all be positive even in the case that $\mathcal{L}$ is Markovian. These coefficients came from the pseudo-Kraus decomposition of $\mathcal{L}$, so if they were all positive, then $\mathcal{L}$ would be a completely positive map. But this need not be the case --- the integral of $\mathcal{L}$ starting from $t=0$ should be completely positive, but $\mathcal{L}$ itself does not need to have this property.

\section{Perturbation theory}
\label{sec:perturbations}

Suppose two finite-dimensional quantum systems $A$ and $B$ interact via a time-independent Hamiltonian $H_{A, \text{bare}} + H_{B, \text{bare}} + \lambda V$, where $V$ acts on both the $A$ and $B$ systems. For convenience, we will use the symbol $H_0$ to denote the combination $H_{A, \text{bare}} + H_{B, \text{bare}}.$ If the initial state of the system is $\rho_A(0) \otimes \rho_B(0)$, then the time evolution of $\rho_A$ is given by
\begin{equation} \label{eq:interacting-channel}
	\rho_A(t)
		= \tr_B\left[
			e^{- i (H_{0} + \lambda V) t}
			(\rho_A(0) \otimes \rho_B(0))
			e^{i (H_{0} + \lambda V) t}
			\right].
\end{equation}
As in section \ref{subsec:superoperator-evolution}, we will package this expression as a quantum channel $\mathcal{N}_{t}$, so we can write
\begin{equation} \label{eq:definition-of-N}
	\rho_{A}(t) = \mathcal{N}_{t}(\rho_A(0)).
\end{equation}
For time scales $t$ that are small with respect to the inverse coupling $1/\lambda$, the dynamics of this system can be computed in $\lambda$-perturbation theory. The time evolution channel $\mathcal{N}_{t}$ can be expanded in $\lambda$, and we will be able to use that expansion to write down a perturbation series for a superoperator $\mathcal{L}_{t}$ satisfying
\begin{equation} \label{eq:target-master-equation}
	\frac{\d \rho_A}{\d t} = \mathcal{L}_{t}(\rho_A)
\end{equation}
to all orders in $\lambda$-perturbation theory.

It is worth pausing to comment on what we mean by ``perturbation theory.'' Since $\mathcal{N}_t$ for any fixed $t$ is formed by exponentials and traces of finite-dimensional matrices, it is analytic in $\lambda.$ It therefore admits a convergent Taylor series of superoperators:
\begin{equation} \label{eq:N-taylor-series}
	\mathcal{N}_{t} = \mathcal{N}_{t}^{(0)} + \lambda \mathcal{N}_{t}^{(1)} + \lambda^2 \mathcal{N}_{t}^{(2)} + \dots.
\end{equation}
Similarly, $\mathcal{L}_{t}$ admits the convergent Taylor series
\begin{equation} \label{eq:L-taylor-series}
	\mathcal{L}_{t} = \mathcal{L}_{t}^{(0)} + \lambda \mathcal{L}_{t}^{(1)} + \lambda^2 \mathcal{L}_{t}^{(2)} + \dots.
\end{equation}
We may also define a convergent $\lambda$-Taylor series for $\rho_A(t)$ by
\begin{equation}
	\rho_{A}^{(k)}(t) = \mathcal{N}^{(k)}_t(\rho_A(0)).
\end{equation}
So we can solve for the superoperator $\mathcal{L}_{t}$ at each order in $\lambda$ via the equation
\begin{equation}
	\frac{1}{k!} \frac{\d^k}{\d \lambda^k} \left. \mathcal{L}_{t}(\rho_A) \right|_{\lambda = 0}
		= \frac{\d \rho_{A}^{(k)}}{\d t}.
\end{equation}
By inserting the Taylor series for $\mathcal{L}_{t}$ and $\rho_A$ into the left-hand side of this equation, we obtain
\begin{equation}
	\sum_{m + n = k} \mathcal{L}_{t}^{(m)}(\rho_A^{(n)}) = \frac{\d \rho_{A}^{(k)}}{\d t}.
\end{equation}
Separating out the $m=k$ terms of this expression gives
\begin{equation}
	\mathcal{L}_{t}^{(k)}(\rho_A^{(0)})
		= \frac{\d \rho_{A}^{(k)}}{\d t} - \sum_{m=0}^{k-1} \mathcal{L}_{t}^{(m)}(\rho_{A}^{(k-m)}).
\end{equation}
Since $\rho_A^{(0)}$ is just $e^{- i H_{A, \text{bare}} t} \rho_{A}(0) e^{i H_{A, \text{bare}} t},$ we may rewrite this as
\begin{equation}
	\mathcal{L}_{t}^{(k)}\left( e^{- i H_{A, \text{bare}} t} \rho_{A}(0) e^{i H_{A, \text{bare}} t} \right)
		= \frac{\d \mathcal{N}_{t}^{(k)}}{\d t}(\rho_A(0)) - \sum_{m=0}^{k-1} (\mathcal{L}_{t}^{(m)} \circ \mathcal{N}_{t}^{(k-m)}) \left( \rho_{A}(0) \right).
\end{equation}
Since $\rho_A(0)$ is an arbitrary unit-trace positive operator, and since this expression is linear in $\rho_A(0)$, it in fact holds for $\rho_A(0)$ any operator. In particular, if $M$ is an arbitrary Hermitian matrix, then setting $\rho_{A}(0) = e^{i H_{A, \text{bare}} t} M e^{- i H_{A, \text{bare}} t}$ yields the following theorem.
\begin{theorem} \label{eq:perturbation-theorem}
	Let $\H_A \otimes \H_B$ be a finite-dimensional Hilbert space with Hamiltonian $H_{A, \text{bare}} + H_{B, \text{bare}} + \lambda V,$ and define $\mathcal{N}_{t}$ and $\mathcal{L}_{t}$ as in equations \eqref{eq:definition-of-N} and \eqref{eq:target-master-equation}. Then $\mathcal{N}_{t}$ and $\mathcal{L}_{t}$ admit convergent Taylor series
	\begin{equation}
		\mathcal{N}_{t} = \sum_k \lambda^k \mathcal{N}_{t}^{(k)}
	\end{equation}
	and
	\begin{equation}
		\mathcal{L}_{t} = \sum_{k} \lambda^k \mathcal{L}_{t}^{(k)}
	\end{equation}
	satisfying
	\begin{equation} \label{eq:recursive-formula}
		\mathcal{L}_{t}^{(k)}\left( M \right)
		= \frac{\d \mathcal{N}_{t}^{(k)}}{\d t}\left( e^{i H_{A, \text{bare}} t} M e^{- i H_{A, \text{bare}} t} \right) - \sum_{m=0}^{k-1} (\mathcal{L}_{t}^{(m)} \circ \mathcal{N}_{t}^{(k-m)}) \left( e^{i H_{A, \text{bare}} t} M e^{- i H_{A, \text{bare}} t} \right).
	\end{equation}
\end{theorem}
\noindent By starting with the first-order term $\mathcal{L}_{t}^{(0)}(M) = - i [H_{A, \text{bare}}, M],$ theorem \ref{eq:perturbation-theorem} can be applied recursively to compute $\mathcal{L}_{t}^{(k)}(M)$ for arbitrarily large $k$ in terms of $\mathcal{N}_{t},$ which in turn can be expressed via equation \eqref{eq:definition-of-N} in terms of $H_{A,\text{bare}} H_{B, \text{bare}}, V,$ and $\rho_B(0).$ In the following, we apply the interaction picture of quantum mechanics to derive explicit formulas for $\mathcal{N}_{t}^{(k)}$ and $\d \mathcal{N}_{t} / \d t.$

Defining $V_{\text{int}}(t)$ by
\begin{equation}
	V_{\text{int}}(t)
		= e^{i H_0 t} V e^{-i H_0 t},
\end{equation}
we know that the time evolution operator can be rewritten as\footnote{This equation is in some sense the fundamental tool in the interaction picture of quantum mechanics. It can be derived by verifying that both sides of the equation satisfy the first-order differential equation $d M/dt = - i (H_0 + \lambda V) M(t)$, and that they agree as operators at $t=0.$}
\begin{equation}
	e^{- i (H_0 + \lambda V) t}
		= e^{- i H_0 t} \left(\mathcal{T} e^{- i \lambda \int_0^t \d t'\, V_{\text{int}}(t')} \right),
\end{equation}
where $\mathcal{T}$ is the time-ordering symbol. Plugging this into equation \eqref{eq:interacting-channel} and expanding in powers of $\lambda$ gives
\begin{align}
	\mathcal{N}_{t}(\rho_A(0))
		= \sum_{k=0}^{\infty} (i \lambda)^k \sum_{m=0}^k \frac{(-1)^{m}}{m!(k-m)!} & 
			\tr_B\left[
				e^{- i H_0 t}
				\left(\mathcal{T} \left( \int_0^t \d t'\, V_{\text{int}}(t') \right)^m \right) \right. \times \nonumber \\ &
				\left. 
				(\rho_A(0) \otimes \rho_B(0))
				\left(\mathcal{T} \left( \int_0^t \d t'\, V_{\text{int}}(t') \right)^{k-m} \right)
				e^{i H_0 t}
		\right].
\end{align}
As a simplification, we can exploit the cyclic property of the trace to cancel the $H_{B, \text{bare}}$ terms appearing in $e^{-i H_0 t}$ and $e^{i H_0 t}.$ This gives the $k$-th order channel $\mathcal{N}_{t}^{(k)}$ as
\begin{align} \label{eq:perturbative-N}
	\mathcal{N}_{t}^{(k)}(M)
		= i^k \sum_{m=0}^k \frac{(-1)^{m}}{m!(k-m)!} &
			\tr_B\left[
			e^{- i H_{A, \text{bare}} t}
			\left(\mathcal{T} \left( \int_0^t \d t'\, V_{\text{int}}(t') \right)^m \right)
			\right. \times \nonumber \\
			& \left. (M \otimes \rho_B(0))
			\left(\mathcal{T} \left( \int_0^t \d t'\, V_{\text{int}}(t') \right)^{k-m} \right)
			e^{i H_{A, \text{bare}} t}
		\right].
\end{align}
Similarly, we can compute the time derivative $\d \rho_A / \d t$ perturbatively, and obtain
\begin{align} \label{eq:perturbative-N-dot}
	\frac{d \mathcal{N}_{t}^{(k)}}{\d t}(M)
	= & - i \left[H_{A, \text{bare}}, \mathcal{N}_{t}^{(k)}(M) \right] - \sum_{m=0}^{k-1} \frac{i^k (-1)^m}{m!(k-m-1)!} \times \nonumber \\
	&
		\tr_B \left[ V, e^{- i H_{A,\text{bare}} t}
			\left(\mathcal{T} \left( \int_0^t \d t'\, V_{\text{int}}(t') \right)^m \right)
			(M \otimes \rho_B(0)) \right. \times \nonumber \\
			& \qquad \left.
			\left(\mathcal{T} \left( \int_0^t \d t'\, V_{\text{int}}(t') \right)^{k-m-1} \right) e^{i H_{A,\text{bare}} t}
		\right].
\end{align}

We now have all the information we need to compute equation \eqref{eq:recursive-formula} to any order we like. The calculation is tedious, but finite. We will report here the explicit computation carried out to quadratic order in $\lambda$, and compute the canonical Hamiltonian to the same order. At leading order, thanks to our definitions, we have
\begin{equation} \label{eq:L-leading}
	\mathcal{L}_{t}^{(0)}(M) = - i [H_{A, \text{bare}}, M],
\end{equation}
for which the canonical Hamiltonian is obviously just $H_{A, \text{bare}}.$ At linear order in $\lambda$, equation \eqref{eq:recursive-formula} gives
\begin{equation}
	\mathcal{L}_{t}^{(1)}(M)
		= \frac{\d \mathcal{N}_{t}^{(1)}}{\d t}(e^{i H_{A, \text{bare}} t} M e^{-i H_{A, \text{bare}} t})
		+ i [H_{A, \text{bare}}, \mathcal{N}_{t}^{(1)}(e^{i H_{A, \text{bare}} t} M e^{-i H_{A, \text{bare}} t})].
\end{equation}
Plugging in formulas \eqref{eq:perturbative-N} and \eqref{eq:perturbative-N-dot} gives the expression
\begin{equation} \label{eq:L-linear}
	\mathcal{L}_{t}^{(1)}(M)
		= - i \left[ \tr_B(V \rho_B(0)), M \right].
\end{equation}
At this order, like at leading order, the canonical Hamiltonian can be extracted by inspection without doing any explicit projections: it is simply $ \tr_B(V \rho_B(0)),$ the partial trace of $V$ (which acts on both the $A$ and $B$ systems) against the initial $B$ state $\rho_B(0).$

At quadratic order in $\lambda$, things get a little more interesting. Using the leading and linear expressions for $\mathcal{L}_t$ given in equations \eqref{eq:L-leading} and \eqref{eq:L-linear}, equation \eqref{eq:recursive-formula} becomes
\begin{align}
	\mathcal{L}_{t}^{(2)}(M)
		= & \frac{\d \mathcal{N}_{t}^{(2)}}{\d t}(e^{i H_{A, \text{bare}} t} M e^{-i H_{A, \text{bare}} t}) \nonumber \\ &
		+ i \left[H_{A,\text{bare}}, \mathcal{N}_{t}^{(2)}(e^{i H_{A, \text{bare}} t} M e^{-i H_{A, \text{bare}} t}) \right] \nonumber \\
		& + i \left[ \tr_B(V \rho_B(0)), \mathcal{N}_{t}^{(1)}(e^{i H_{A, \text{bare}} t} M e^{-i H_{A, \text{bare}} t}) \right].
\end{align}
Plugging in formulas \eqref{eq:perturbative-N} and \eqref{eq:perturbative-N-dot} and simplifying gives
\begin{align}
	\mathcal{L}_{t}^{(2)}(M)
	= & \tr_B \left[ V - \tr_B(V \rho_B(0)) \otimes I_B, \left[M \otimes \rho_B(0),
	e^{- i H_{A,\text{bare}} t} \left( \int_0^t \d t'\, V_{\text{int}}(t') \right) e^{i H_{A,\text{bare}} t}\right]
	\right].
\end{align}
We can compute the canonical Hamiltonian for this superoperator using the tensor network equation \eqref{eq:canonical-ham-TN} from section \ref{subsec:projection-computation}. It will be helpful to simplify notation a bit. Let us write $\tilde{V}(t)$ for $e^{- i H_{A, \text{bare}} t} \int_0^t \d t' V_{\text{int}}(t') e^{i H_{A, \text{bare}} t}$, and $\hat{V}$ for $\tr_B(V \rho_B(0)).$ Under these notational substitutions, the quadratic contribution to the master equation becomes
\begin{align}
	\mathcal{L}_{t}^{(2)}(M)
	= & \tr_B \left[ V - \hat{V} \otimes I_B, \left[M \otimes \rho_B(0),
	\tilde{V}(t) \right]
	\right].
\end{align}
We can write the superoperator $\mathcal{L}^{(2)}_t$ in tensor network notation as the following, fairly lengthy expression, in which we have used red lines to denote legs belonging to the $B$ system and black lines to denote legs belonging to the $A$ system.
\begin{equation}
	\includegraphics{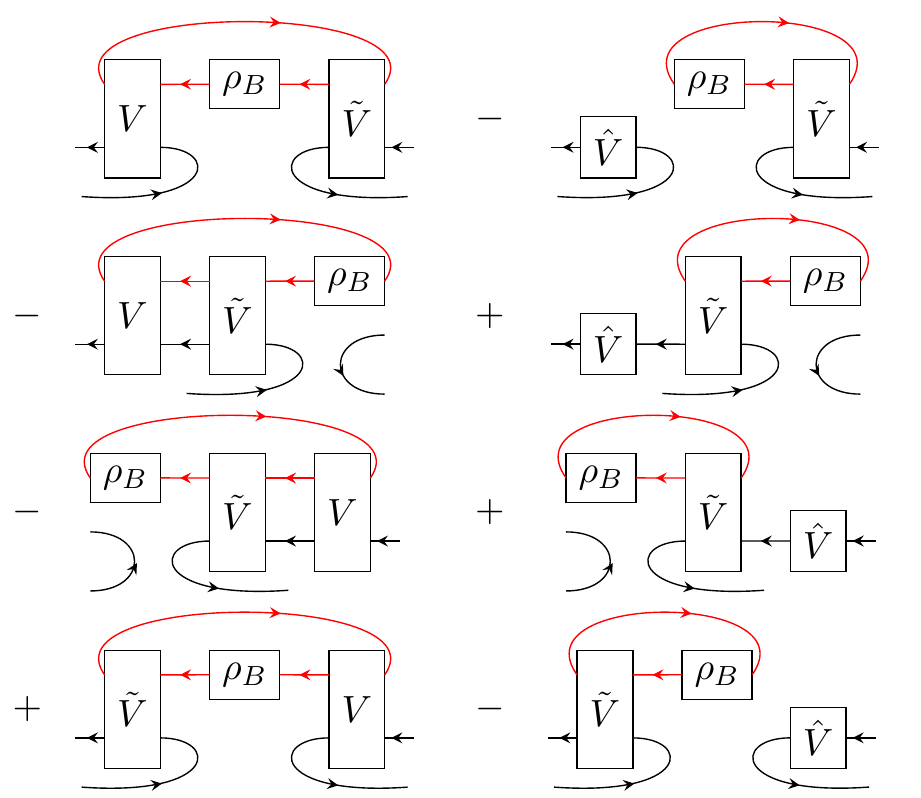}
\end{equation}

All that remains is to plug into equation \eqref{eq:canonical-ham-TN} and simplify. We omit the calculation, and report only the final result. Making use of the notational shorthands $\tilde{V}_B(t) = \tr_A(V(t))$ and $V_B = \tr_A(V)$, the final answer is
\begin{align} \label{eq:quadratic-canonical-hamiltonian}
	H_A^{(2)}
		= & \frac{1}{2 i d}
			\tr_B\left(\tilde{V}(t) [V_B, \rho_B(0))]\right)
			+ \frac{1}{2 i d} \tr_B\left( V[\tilde{V}_B(t), \rho_B(0)]\right)
				\nonumber \\
			& + \frac{1}{2i}
				\tr_B\left([V, \tilde{V}(t)] \rho_B(0) \right)
				+ \frac{1}{2i} \left[ \tr_B(\tilde{V}(t) \rho_B(0)), \hat{V} \right]
				\nonumber \\
			& + \alpha I_A,
\end{align}
where we have packaged all terms proportional to the identity into a single coefficient $\alpha$ --- these do not affect the dynamics, and are fixed only by our convention that the canonical Hamiltonian be traceless. A few comments are in order. The first is that the canonical Hamiltonian generically becomes time dependent at quadratic order in $\lambda$, while the bare Hamiltonian $H_{A, \text{bare}}$ and the linear correction $H_A^{(1)} = \tr_B(V \rho_B(0))$ were time-independent. The second is that if we formally take the limit $d\rightarrow \infty$, the first two terms in the quadratic canonical Hamiltonian disappear. In reality, one must be careful in taking such a limit --- the operators $V_B$ and $\tilde{V}_B(t)$, since they are defined using partial traces over the $A$ system, may scale with $d$, and so depending on how one takes the limit those terms might not truly vanish. If we momentarily throw caution to the wind and clothe our equal signs in scare quotes, however, then we obtain 
\begin{align}  \label{eq:infinite-d-Hamiltonian}
	\lim_{d \rightarrow \infty}
	H_A^{(2)}
	 \text{``}=\text{''} \frac{1}{2i}
	\tr_B\left([V, \tilde{V}(t)] \rho_B(0) \right)
	+ \frac{1}{2i} \left[ \tr_B(\tilde{V}(t) \rho_B(0)), \hat{V} \right]
	+ \alpha I_A.
\end{align}
If we fix $\rho_B(0)$ to be a pure eigenstate of $H_{B, \text{bare}}$, labeled by $\rho_B(0) = \ketbra{\bar{u}}{\bar{u}}$, then it is straightforward to check that equation \eqref{eq:infinite-d-Hamiltonian} agrees, up to the dynamically unimportant trace term, with equation (24) from \cite{agon2018coarse}. That equation describes one possible choice of effective Hamiltonian for the IR-UV open system dynamics of a quantum field theory. While our equation \eqref{eq:quadratic-canonical-hamiltonian} implies that their chosen Hamiltonian is \emph{not} the optimal Hamiltonian for finite-dimensional systems, the formal limit in equation \eqref{eq:infinite-d-Hamiltonian} suggests that their chosen Hamiltonian may be in some sense the right canonical choice in infinite dimensions. We leave the problem of precisely defining the canonical Hamiltonian in infinite dimensions to future work.

\acknowledgments{We thank Dan Ranard and Amir Safavi-Naeini for enlightening conversations. We especially thank Dan Ranard for teaching us tricks for manipulating tensor network equations that helped simplify the presentation of section \ref{subsec:projection-computation}. The authors are supported by AFOSR award FA9550-19-1-0369, CIFAR, DOE award DE-SC0019380 and the Simons Foundation.}

\bibliographystyle{JHEP}
\bibliography{biblio}

\end{document}